# Data-driven Analysis of Regional Capacity Factors in a Large-Scale Power Market: A Perspective from Market Participants

Zhongyang Zhao, *Student Member, IEEE*, Caisheng Wang, *Senior Member, IEEE*
*Department of Electrical and Computer Engineering, Wayne State University, Detroit, MI, USA*
Huaiwei Liao, *Member, IEEE*
Carol J. Miller
*Department of Civil and Environmental Engineering, Wayne State University, Detroit, USA*

***Abstract*— A competitive wholesale electricity market consists of thousands of interacting market participants. Driven by the variations of fuels costs, system loads and weathers, these market participants compete actively and behave variously in the power market. Although electricity markets tend to become more transparent, a large amount of market information is still not publicly available to market participants. Hence, data-driven analysis based on public data is crucial for market participants to better understand and model large-scale power markets, and ultimately to perform better in power trading. While most of the previous researches related to the large-scale power markets are based on the synthetic networks, a data-driven approach utilizing the real power market data is proposed in this paper. First, the power plants' monthly net generation and capacity data are obtained from U.S. Energy Information Administration (EIA) and aggregated to figure out the monthly regional capacity factors which are used to characterize the market's regional behaviors for market participants. Then, the regional capacity factors are analyzed against the metered system loads and natural gas prices to study the generation behaviors in the power market. The analysis reveals the impacts of regional natural gas prices on capacity factors and the responses of generating behaviors to the system loads. The analysis results present the solid evidence and rational references for market participants to model and validate the large-scale power market in the future.**

## I. INTRODUCTION

In United States, there are more than 7,300 power plants, nearly 160,000 miles of high-voltage power lines and millions of low-voltage power lines and distribution transformers in the power grid serving 145 million customers [1]. To efficiently and economically manage large power grid for their safe and reliable operations, wholesale and retail markets have been formed throughout the country. There are several competitive wholesale markets in United States run by Independent System Operators (ISO) or Regional Transmission Organizations (RTOs), such as Midcontinent Independent System Operator (MISO), California ISO (CAISO), Electric Reliability Council of Texas (ERCOT), New York ISO, New England ISO and PJM Interconnection LLC (PJM). Among these market-based power systems, the PJM market is the largest. It has all or parts of 13 states and the District of Columbia under regulation and serves about 65 millions of people [2]. With 21% of the U.S. Gross Domestic Product (GDP) produced in its region, the PJM's summer peak load can hit 165.49 GW with a generation capacity of 178.56 GW [2].

Since lots of information in the large-scale power market, including detailed generator offer data and fuel costs, is not publicly available to market participants, it is difficult to study the market's behaviors in order to better understand and model the power markets. To overcome the difficulties due to the lack of information, several methodologies have been proposed to create large-scale synthetic network test cases for matching realistic structural and statistical characteristics based on the clustering techniques [3], [4]. Furthermore, economic studies based on a large-scale synthetic network were carried out to determine the generator cost models [5]. In addition to modeling power plants' heat rates, fuel types, and cost models, the fuel costs analytics and predictions for validating the large market were pursued as well [6]. Even though statistical analyses have been performed for characterizing large-scale power markets, the data-driven analysis based on the real power market data towards the generating behaviors has been barely studied. As aforesaid, there are spacious areas from Midwest to east coast in PJM market. These areas across Appalachian Mountains have various shale gas basins, coal mines, natural gas (NG) pricing hubs, transmission limits, weather conditions, populations, states policies, etc. These factors can significantly affect or even determine power plants' generation behaviors over not only spatial but also temporal domains. Therefore, analyzing various market generation behaviors over the extensive area based on the public market data will be beneficial for market participants and researchers to comprehend and model a large-scale power market more accurately.

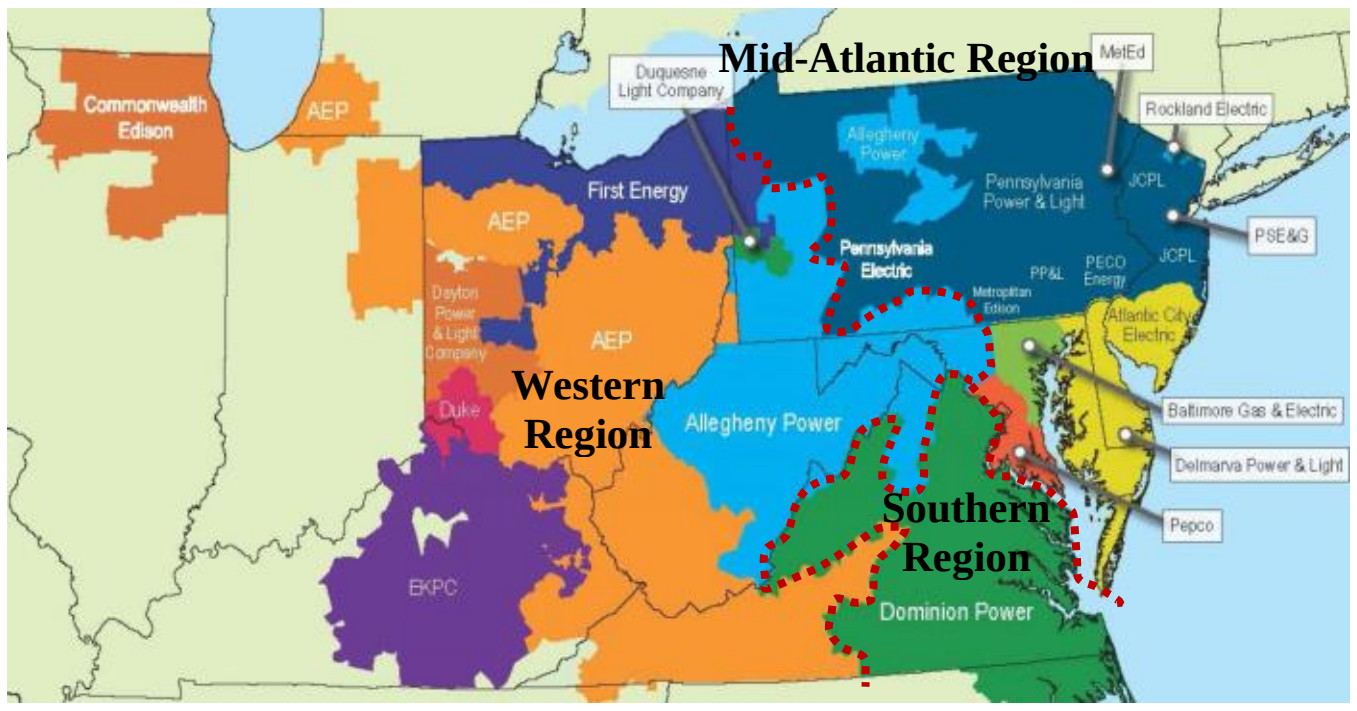

Fig. 1. PJM map with regions [7], [8].

In this paper, the monthly regional capacity factor (RCF), referring to the ratio of the net generation to the corresponding generation capacity over a month in a region, is used to characterize the market's regional behavior. The advantage of monthly RCF is to capture the characteristics from a group of same-region plants, which provides a good opportunity to discover the regional patterns of market behaviors. In PJM, three regions are aggregated from the load zones [2], [7] as shown on the PJM map in Fig. 1. The dashed red lines on the map split the areas of Western Region, Mid-Atlantic Region and Southern Region in PJM. This paper focuses on the PJM market and applies the RCFs on these regions to analyze the collective generating behaviors.

To study the monthly RCFs in PJM, this paper first explores the monthly data of EIA-923 [9] and EIA-860M [10] and proposes the criteria for selecting the objective power plants located in the PJM regions for calculating the RCFs. Then, the NG pricing hubs in PJM and the NG prices collected from EIA Open Data [11] are introduced for the following RCF study. In addition to the NG prices, the system loads obtained from PJM Data Miner 2 [12] are used for supporting the analytics as well. The results of the data-driven analysis show that the proposed RCF method is effective to observe and distinguish the seasonal patterns of generating behaviors in different regions. Meanwhile, the extracted NG prices and the load information are capable of further illustrating the observed phenomena from the RCFs.

## II. Featured Data

The data for obtaining RCFs are presented and preprocessed in this section. The raw data regarding to the monthly net generation and capacity of the power stations by different fuel types are collected from Forms EIA-923 and EIA-860M, respectively. Because the earliest monthly EIA-860M data is July 2015, the monthly generation and capacity data from July 2015 to December 2017 are collected and utilized to calculate the RCFs. To further study the RCFs, the regional NG fuel costs are obtained according to the NG pricing hubs information provided by PJM. Meanwhile, the monthly system-level loads are lumped together from the hourly load data in PJM.

### A. Power Plants Monthly Data

In Form EIA-923, the survey collects the electricity generation and fuel consumption of various power plants [9]. Similarly, Form EIA-860M reports electric power generation capacity at the unit level [10]. The public information about the power plant's monthly net generation and capacity extracted from these two EIA forms are preprocessed and aggregated into plant-level by different fuel types to prepare for the regional capacity factor investigation in PJM.

Table I Monthly Data of Forms EIA-923 and EIA-860

| Date | Plant ID | Plant State | Energy Source | Net Generation (MWh) | Capacity (MW) |
|---|---|---|---|---|---|
| 2015-07 | 1554 | MD | Coal | 158687 | 423 |
| 2015-07 | 1554 | MD | NG | 10096 | 126 |
| 2015-08 | 1554 | MD | Coal | 47207 | 423 |
| 2015-08 | 1554 | MD | NG | 3343 | 126 |

Part of the aggregated data are listed in Table I for reference. Table I shows the monthly net generation and generation capacity of a power station located in Maryland. These kinds of preprocessed information are employed for the following analysis.

As introduced in the previous section, there are 14 areas governed in PJM market and these areas are separated into Western Region, Mid-Atlantic Region, and Southern Region. Because the Western Region and Mid-Atlantic Region own over 80% of the generation capacity of PJM while Southern Region has the rest [10], the power plants located in the two larger regions are selected for studying the regional capacity factors, respectively.

Furthermore, compared to larger power stations (i.e., larger than 200 MW in generation capacity), there are missing data issues for smaller power stations in Form EIA-923. Meanwhile, 89.26% generation capacity of PJM is contributed by the power plants with over 200 MW capacity [10], which means the generating behaviors of these large power plants can sufficiently represent the overall market behaviors.

Therefore, the reported monthly generation and capacity data from July 2015 to December 2017 of the power plants located in either Western Region or Mid-Atlantic Region, which have a generation capacity over 200 MW, are selected for the further regional capacity factor analysis, and the criteria for selecting the objective power plants to analyze the monthly capacity factor of a region are displayed in Fig. 2. The central common area represents the objective power plants chosen for the RCF analysis in this paper.

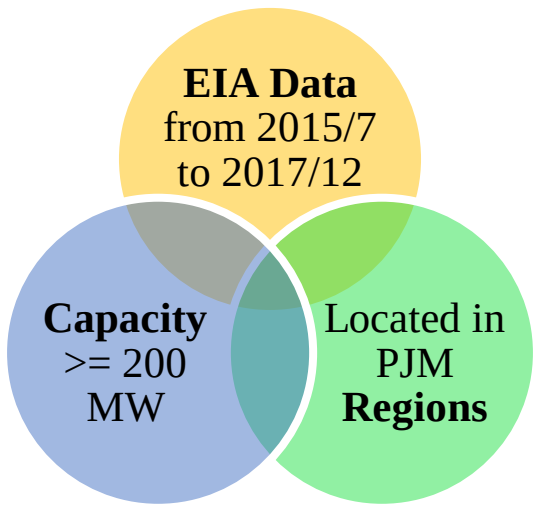

Fig. 2. Proposed criteria to select power plants.

At this point, 112 power stations are chosen for representing the Western Region's power plants and 104 power stations are picked for the Mid-Atlantic Region. By aggregating the selective power plants' capacities into each region, a total of 85.49 GW and 68.28 GW generation capacities are obtained for the Western Region and Mid-Atlantic Region, respectively. As the capacity bar plots of different fuel types shown in Fig. 3, 26.97 GW capacity is contributed by NG and 16.64 GW is Nuclear in Western Region. Additionally, the coal-fired generation owns 37.75 GW while the rest are generated by other units including wind, hydro, etc. In Mid-Atlantic Region, 29.82 GW, 17.64 GW, 13.80 GW and 7.01 GW capacities are provided by NG, coal, nuclear and other generation units correspondingly. The major difference between these two regions is the coal capacity in Western Region beats the coal capacity of Mid-Atlantic Region by 20.11 GW.

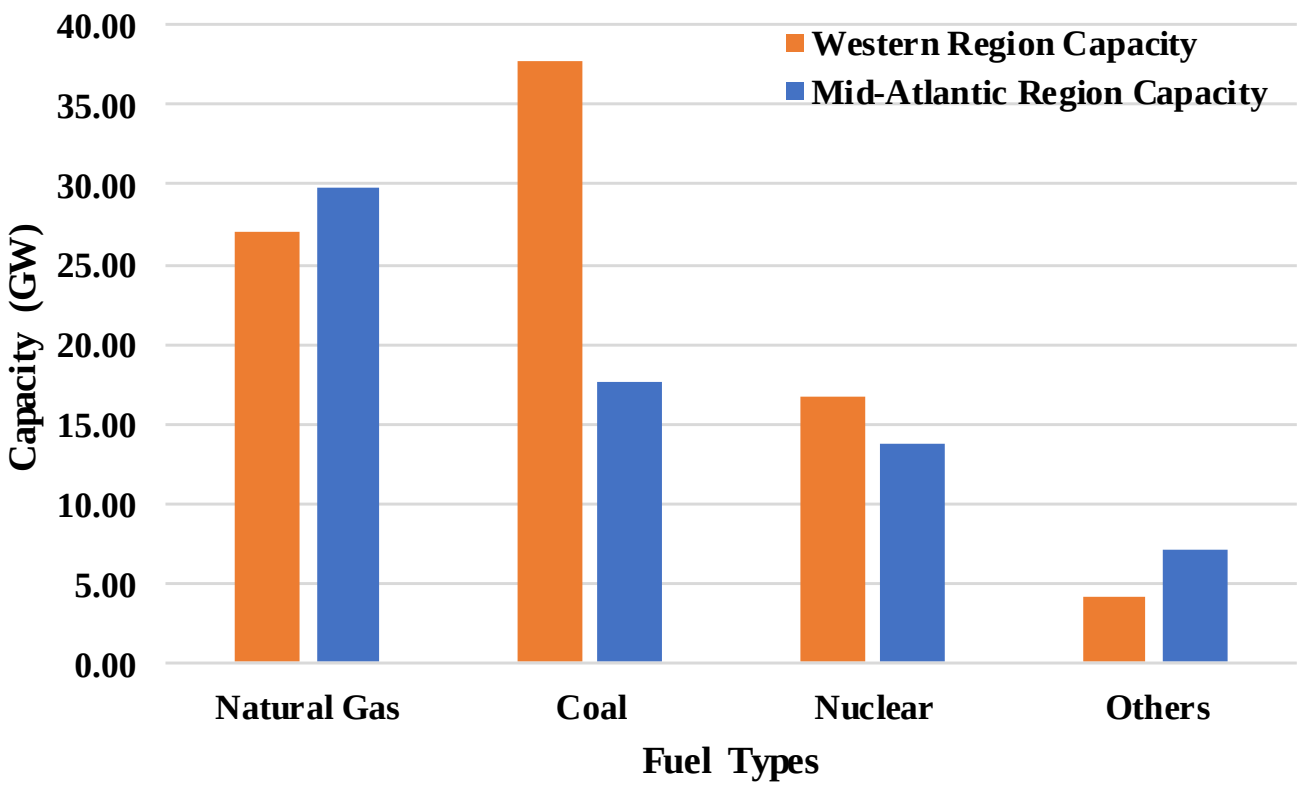

Fig. 3. Capacity by fuel types in the PJM regions.

Then, utilizing the collected data of the power plants in Western Region and Mid-Atlantic region, the regional capacity factor can be calculated by (1):

$$RCF^M = \frac{\sum_i^n Plant\ Netgen_i^M}{\sum_i^n Plant\ Capacity_i^M \times Hours^M} \quad (1)$$

where $Plant\ Netgen_i^M$ represents the generated energy of plant $i$ in month $M$; $Plant\ Capacity_i^M$ stands for the plant-level monthly generation capacity; $Hours^M$ means the total generating hours in month $M$; $n$ is the number of the power plants that are selected for evaluating the regional capacity factor $RCF^M$. The obtained $RCF^M$ is employed to investigate the generating behaviors in the large-scale market.

## B. *Natural Gas Fuel Cost*

According to the capacity bar plots of different fuel types in Fig. 3, a large portion, close to 70%, of the regional generation capacity is from NG and coal in the both target PJM regions. For the sake of analyzing the RCFs and better studying the market behaviors, the NG fuel cost naturally becomes an essential factor that needed to be addressed. Since the fuel cost of coal has much less volatility than the NG price [6], the coal cost is not further discussed for the RCF study in this paper.

Meanwhile, the PJM spreads over 14 areas from Midwest to east coast, in which the natural gas fuel costs vary significantly among these areas [13]. The fuel cost variances due to the weather change, locational congestion of the NG pipeline, etc., are able to cause the power stations behave differently, which means the variation of fuel costs of different regions in PJM has considerable impacts on the RCFs. The analysis of the RCFs associated with the regional natural gas fuel costs can help market participants and researchers to have an insight into the larger-scale power market.

Table II REGIONAL NATURAL GAS PRICES

| Date | Western Region NG Price ($/MMBtu) | | | | Mid-Atlantic Region NG price ($/MMBtu) | | | |
|---|---|---|---|---|---|---|---|---|
| | IL | MI | OH | Avg. | PA | NJ | NY | Avg. |
| 2017-01 | 3.95 | 3.63 | 3.84 | 3.80 | 4.12 | 4.06 | 5.41 | 4.53 |
| 2017-02 | 3.56 | 3.18 | 3.41 | 3.38 | 3.21 | 3.64 | 5.48 | 4.11 |
| 2017-03 | 4.06 | 3.16 | 2.33 | 3.18 | 2.86 | 3.45 | 2.95 | 3.08 |

According to the information collected from PJM Natural Gas Hub Recommendation [13] and 2018 PJM Energy Market Offer Caps Report [14] published by Monitor Analytics [15], the NG pricing hubs associated with Mid-Atlantics Region are recognized as Transco Zone 6, TETCO M-3 and so on while Western Region NG prices are most affected by Chicago Citygates, Michigan Consolidated, and Dominion-South.

However, the monthly data of these pricing hubs are not openly accessible. Instead, the monthly price for the NG sold to electric power consumers by areas can be collected from EIA Open Data [11] and aggregated to the regional NG fuel costs. Therefore, based on the corresponding areas of the discovered NG pricing hubs, the NG fuel costs data of Illinois, Michigan and Ohio obtained from the EIA are aggregated for characterizing Western Region NG prices. The NG prices data of Pennsylvania, New Jersey and New York are combined for describing Mid-Atlantic Region NG prices. The monthly NG prices data of each preferred state from January 2017 to March 2017 are listed in Table II for reference. The monthly average prices of Western Region and Mid-Atlantic Region are used for supporting the analytics of RCFs in the next section.

## C. *System Load*

In addition to the NG prices, the system-level load is another important factor for better understanding the power market. The power demands during the summer and winter are much higher than the demands along with the shoulder months, which implies the RCFs could vary along with seasonally changing power demands. Hence, it is beneficial to incorporate with the system load information to find out the characteristics and patterns of the RCFs in the power market.

From PJM Data Miner 2, the hourly system-level loads can be procured. For the sake of getting clearer pattern of the monthly load, the hourly load during peak hours [16] from 7:00 to 23:00 on each day are selected and analyzed. In other words, the loads during the off-peak hours which have less information to represent the patterns of monthly loads are removed. Therefore, the monthly loads can be obtained by the process (2):

$$Sytem\ Load^M = \frac{\sum_{d=1}^{Days^M} \sum_{h=7}^{23} Hourly\ Load_{h,d}^M}{Hours^D \times Days^M} \quad (2)$$

where $Hourly\ Load_{h,d}^M$ represents the hourly load at hour $h$ of day $d$; $Hours^D$ means the hours of each day, which is always equal to 16 for the length of peak hours in this paper; $Days^M$ serves as the total days in month $M$ and $Sytem\ Load^M$ is the processed monthly load.

# III. DATA-DRIVEN ANALYSIS OF REGIONAL CAPACITY FACTOR

An in-depth analysis based on the gathered regional capacity factors, regional NG prices and the system-level load from July 2015 to December 2017 is conducted in this section.

## A. *Regional Capacity Factors*

Via selecting the power plants data based on the criteria shown in Fig. 2, the power plants monthly output and capacity can be aggregated to monthly RCFs by (1). The monthly RCFs are plotted with the monthly system-level power demands obtained from PJM in Fig. 4 for comparison and verification.

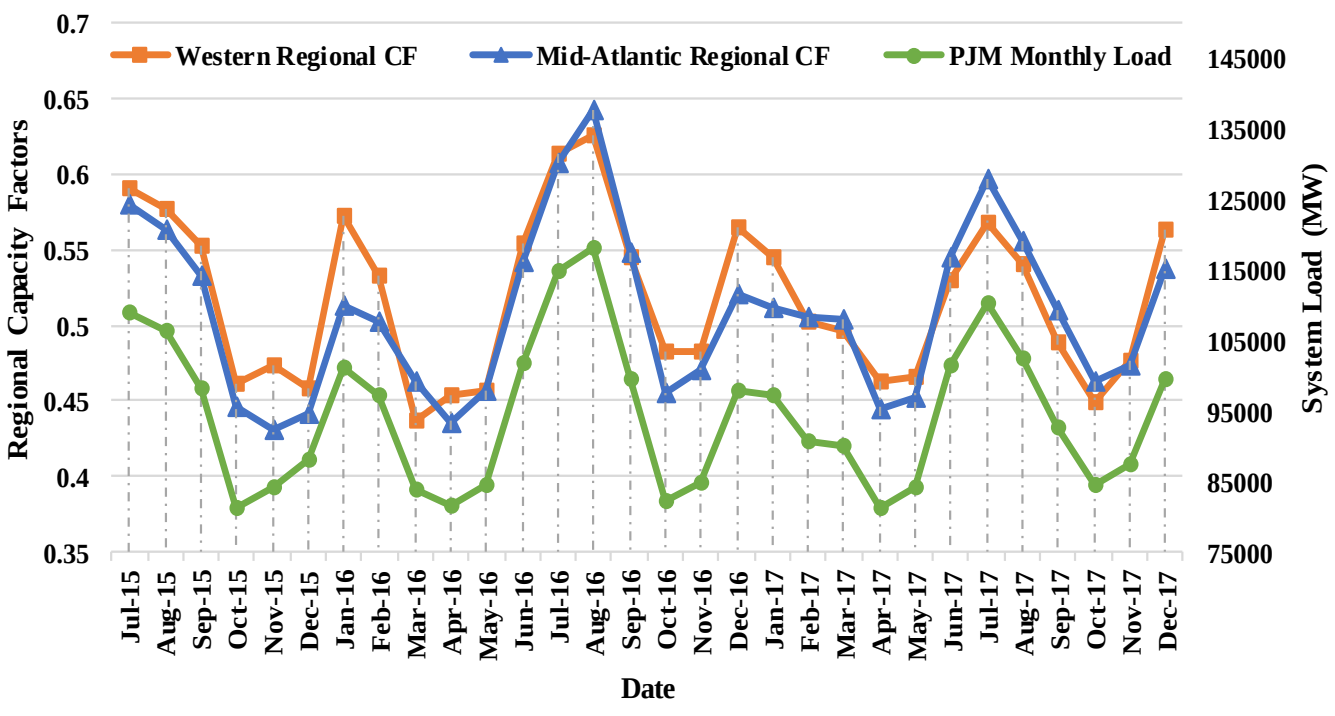

Fig. 4. Series plot of regional capacity factors and system load.

In Fig. 4, the vertical axis on the left side indicates the values of RCFs while the orange and blue solid lines represent the Western and Mid-Atlantics regional capacity factors, respectively. Another vertical axis on the right side shows the values of the system load displayed by the solid green line.

By observing the 30-month variation tendencies of the RCFs, the strong seasonal patterns are clearly identified. While comparing the profiles between the plots of RCFs and monthly system demands, they are almost identical throughout the 30-month data. On the one hand, the cause of these similar profiles can be that the variation of system power demand drives the seasonal changes in the regional generation. On the other hand, the strong correlations between the load and the RCFs verify that the proposed criteria shown in Fig. 2 for selecting the objective power plants data perform well and the selected monthly net generation and capacity can truly and sufficiently represent the regional generation behaviors in PJM. While the RCF data and load data are collected from data sources of EIA and PJM separately, this matching outcome creates a more confident and convincing condition for the following data-driven study.

Although the periodic peaks of the RCFs driven by the system load happen during the summer months from June to September and the winter months between December and February, their characteristics are obviously different in summer and winter. During the winter months, the RCFs in the Western Region are evidently higher than those in the Mid-Atlantic Region. In the summer, the situation is almost flipped over. Both regions' capacity factors come to a similar level and Mid-Atlantic RCFs can even go over Western RCFs in some months, such as August 2016, July 2017 and August 2017.

For the distinction between summer and winter peaks, the system load shown in Fig. 4 does not have sufficient information to support any further analysis. In the followings, the regional NG prices are used to analyze the different characteristics of the summer and winter peaks.

## B. *Natural Gas Fuel Cost Impacts*

As introduced in Section II, the NG fuel cost is an nonneglible factor for analyzing the power staion's generating behaviors. Especially when the development of the shale gas techniques has been driving the overall NG prices down and stable in the last five years, the marginal costs of the coal-fired and NG-fired power plants come to a similar level. Once these two kinds of power stations have similar marginal cost, they become the closest competitors in the power market and the volatility of the NG prices in different seasons will definitely influence the generation of not only the NG-fired power plants but also the coal-fired power plants.

To study the impacts of NG prices on the RCFs, the aggregated regional NG prices from July 2015 to December 2017 are plotted with RCFs in Fig. 5. In this figure, the vertical axis on the left side shows the values corresponding to RCFs and the other vertical axis on the right side represents the NG monthly prices.

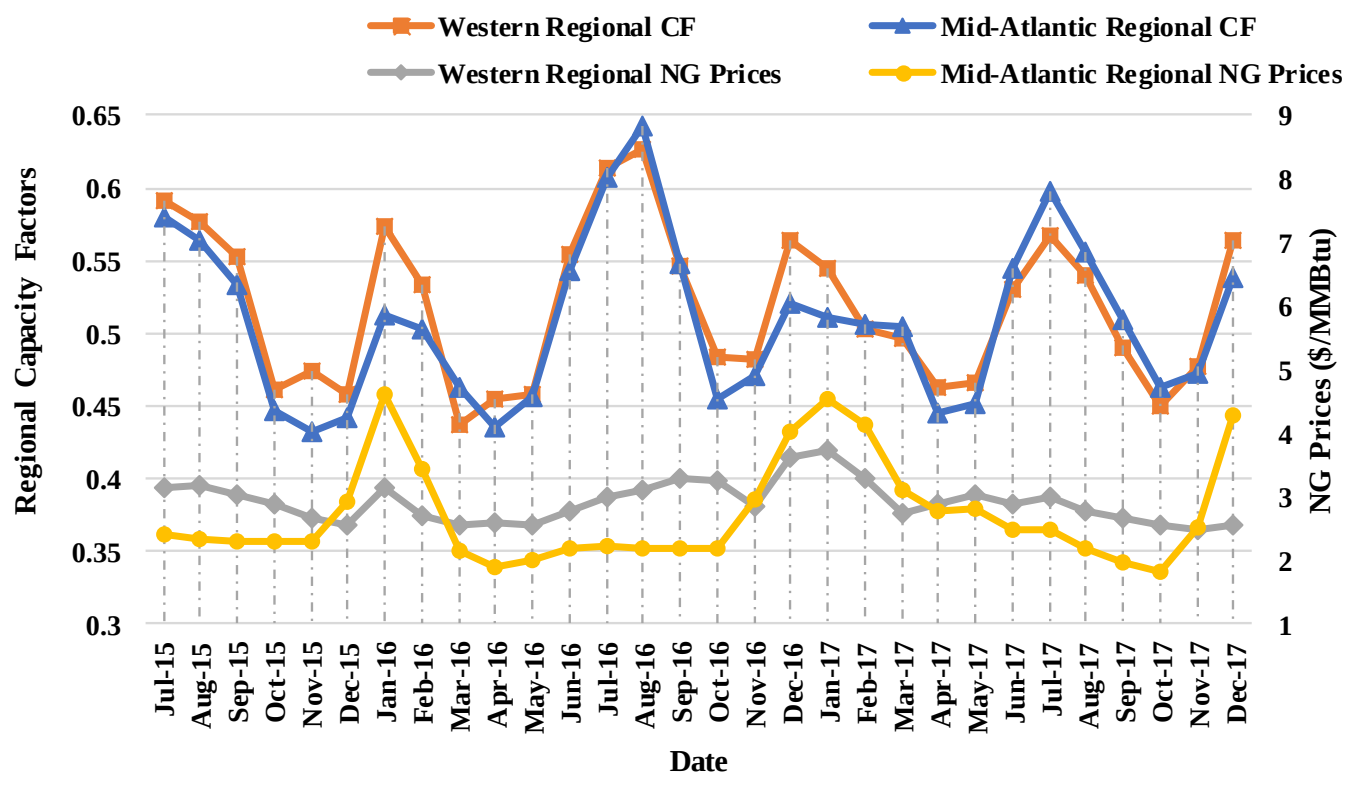

Fig. 5. Series plot of regional capacity factors and regional NG prices.

Comparing the monthly RCFs along with the regional NG prices, we find the reason for the different characteristics among the RCF summer peaks and winter peaks. While the Mid-Atlantic Region NG prices rise during the winter months, the competitiveness of Mid-Atlantic NG-fired power plants in the power market is weakened. At this situation, the Western NG-fired power plants have more advantages to compete in the market, which leads Western RCF to surpass Mid-Atlantic RCF in winter.

While the seasonal variations of the NG prices are discovered to have significant impacts on the RCFs, the power plants that are most influenced by NG prices should be further analyzed. Therefore, the monthly generation and capacity of coal and NG are extracted from the selected power plants data and utilized for calculating the NG and coal RCFs via (1).

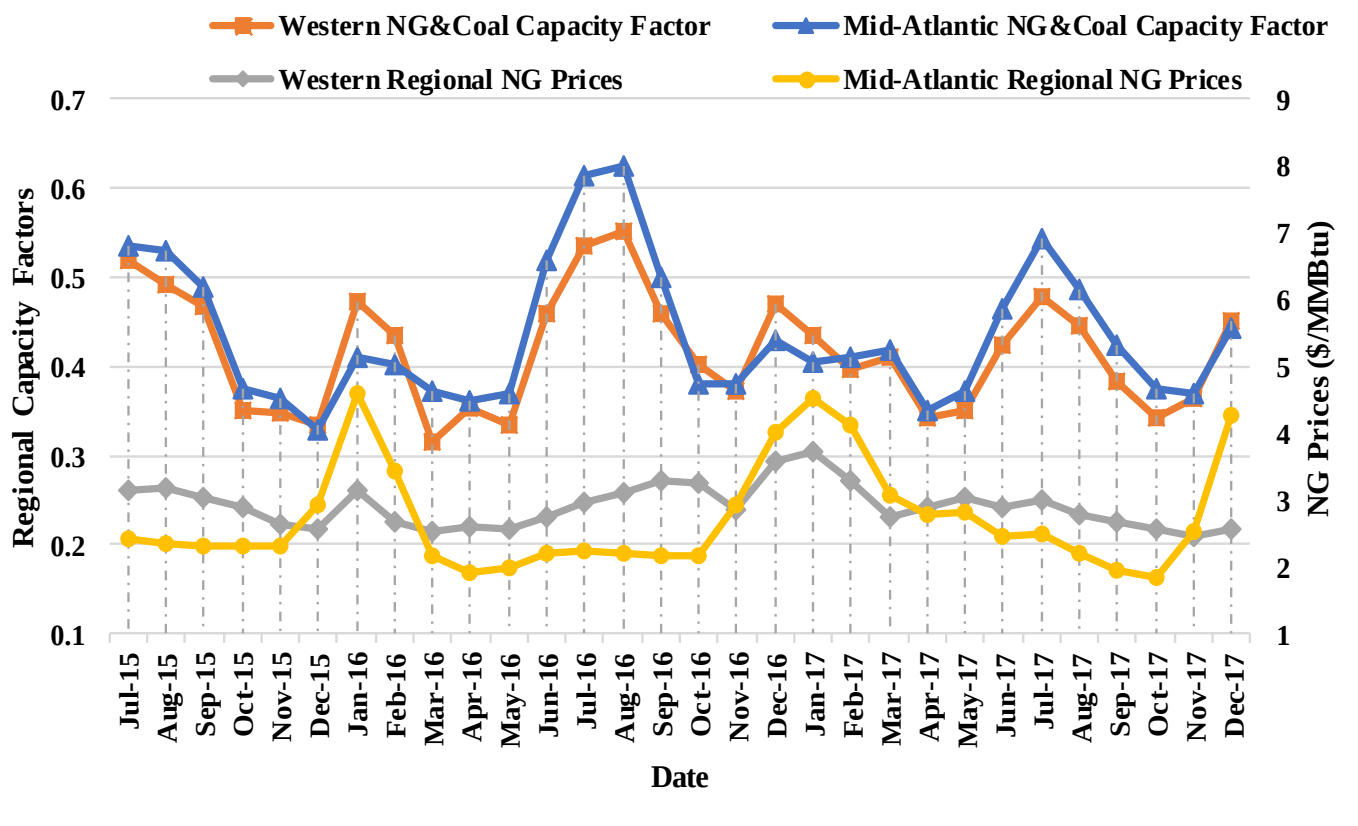

Fig. 6. Series plot of NG and coal regional capacity factors and regional NG prices.

To further address the impacts of NG prices, the monthly NG and coal RCFs are plotted with regional NG prices in Fig. 6 as well. In addition to the previously observed distinctions among the winter RCFs, the RCFs of Mid-Atlantic are found to be substantially larger than Western NG and coal RCFs during the non-winter months when the Mid-Atlantic NG price plunges to a lower level than Western NG price. Eventually, the different characteristics of RCFs between the winter and non-winter months are revealed to be caused by the seasonal variation of the regional NG fuel costs.

### C. *Regional Capacity Factors Responds to System Load*

Even though the seasonal differences between the Western and Mid-Atlantic RCFs are not able to be illustrated by system load in Fig.4, the relaitons between the RCFs and the system load are important to be analyzed as well. In other words, the responses of the regional power plants to the system-level load are crucial to be analyzed for comprehensively understanding their behaviors in a large-scale power market.

In contrast to the nuclear power statoins, the NG and coal power plants reacts to the load changes more sensitively. Considering the seasonality of NG prices, the monthly load and RCFs of NG and coal are extracted and separated into two sets. One set is corresponding to the non-winter months from March to November and the other one is for the winter season containing December, January and February.

These two data sets are plotted separately in Fig. 7, where the vertical axis indicates the values of RCFs and horizontal axis shows the levels of system demands. Futhermore, for the purposes of clearer observations and better comparisons, the linear regressions of the RCFs versus system loads are shown in Fig. 7 as well. In the plot of non-winter data set, the slope of linear regression for Mid-Atlantic Region is larger than Western Region's slope. In the plot of winter data set, the relationship between the two regions is opposite to that in non-winter months. Besides, both plots show that the differences between the RCFs get bigger when the system loads increase to a higher level.

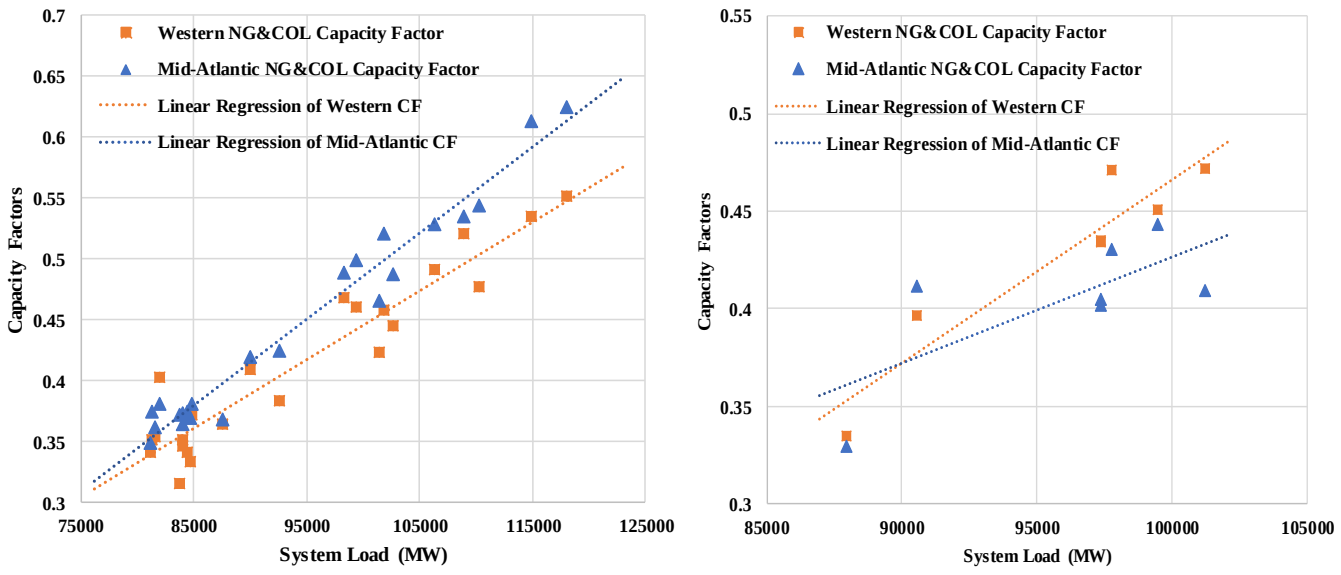

Fig. 7. Scatter plots of regional capacity factors versus system loads in non-winter months (left) and winter months (right).

Overall, it is clearly discovered that the NG prices have the capability to lead power stations responding differently to the various system load levels. Moreover, recalling the higher coal capacity of Western Region shown in Fig. 3, the corresponding sensitivities of RCFs to system loads indicate there are more greenhouse gases emissions in the winter when the load increases.

## IV. CONCLUSION

In this paper, a data-driven approach based on the real power market data has been proposed for analyzing the power market collective behaviors, which helps the market participants to better understand and model the large-scale power market. The RCF index and the criteria for selecting power plants to calculate the index have been developed. The PJM is taken as an example of applying the proposed RCF approach for analyzing the regional generating behaviors. The correlations among the RCFs, the system loads and NG prices were analyzed for the two PJM regions, i.e., Western Region and Mid-Atlantic Region.

The analysis results show that the RCFs have a clear and strong relationship with the system loads and NG prices in the region. The results also verify the effectiveness of the proposed RCF index in characterizing the generation behaviors in a large-scale power market while the power plant selection criteria are based on the real market data. The relationship between the RCFs and NG fuel cost also suggests that not only the distinctions on spatial domain but also the variations on the temporal domain should be considered to characterize a large-scale power market. Furthermore, the higher Western RCFs are noted in the winter, which is highly possible to cause the transmission congestions from Western Region to Mid-Atlantic Region in the PJM market.